\documentclass[
    ,final            
  ]
  {aipproc}

\layoutstyle{6x9}

\begin{document}

\title{Production of Prompt Photons}


\author{Edmond L. Berger}{
  address={High Energy Physics Division, Argonne National Laboratory, Argonne, IL, 60439}
}

\begin{abstract}
After a brief review of the production dynamics of prompt single 
photons in hadron collisions, I summarize a new QCD calculation of the 
transverse momentum distribution of continuum prompt photon pairs 
produced by QCD subprocesses, including all-orders soft-gluon resummation 
valid at next-to-next-to-leading logarithmic accuracy.  Resummation is 
necessary to obtain reliable predictions, as well as 
good agreement with data from the Fermilab Tevatron, in the range of 
transverse momentum where the cross section is largest.  Predictions are 
made for the Large Hadron Collider where the QCD diphoton continuum is 
shown to have a softer 
spectrum in transverse momentum than the Higgs boson signal.
\end{abstract}

\maketitle

\section{Introduction}
The study of {\em prompt} photons carrying large values of transverse 
momentum has a history almost as old as quantum chromodynamics (QCD).  It 
ranks with deep-inelastic-lepton scattering, massive-lepton-pair production 
(the `Drell-Yan process'), and jet production as an important probe of 
short-distance hadron dynamics\footnote{A photon is said to be prompt if 
it cannot be said to originate from the decay of a hadron, such as a 
$\pi^0$ or $\eta$, itself produced with large transverse momentum.}.
The appeal of prompt photons is that they are point-like, colorless probes  
of the dynamics of quarks and gluons, ones that escape unscathed through 
the colored medium of the high-energy collision. One motivation, in 
spin-dependent and spin-averaged scattering, is the access 
that prompt photons offer to the spin-dependent and spin-averaged 
gluon densities of hadrons.  One 
of the two leading-order partonic direct subprocesses feeds directly 
from the gluon parton density through the gluon `Compton' subprocess, 
$q g \rightarrow \gamma q$. 

In reality, the promise and naive simplicity of prompt photons are 
compromised by two issues: fragmentation contributions and photon isolation.  
Fragmentation is a long-distance process in which a hard photon brems off 
a final-state quark (or gluon).  The photon emerges as part of a jet if the 
opening 
angle between the quark and photon is too small.  This `showering' process 
must be parameterized by a non-perturbative single-photon fragmentation 
function $D_{\gamma}(z, \mu_f)$.  Variable $z$ is the fraction of the parent  
parton's momentum that the photon retains, and fragmentation scale $\mu_f$ 
serves as a boundary in momentum $\mu$ space between the long-distance dynamics 
included in $D_{\gamma}(z, \mu_f)$ and the short-distance next-to-leading 
order dynamics in the region $\mu > \mu_f$ (i.e., large angle separation between 
the final photon and remnants of the fragmenting parton).  Note that a new 
variable $\mu_f$ enters the mix along with the usual renormalization and 
factorization scales $\mu_R$ and $\mu_F$.  

A theory calculation is most reliable for an {\em inclusive} observable since 
infra-red and collinear singularities can be shown to cancel between the 
real-emission and loop diagrams in an inclusive calculation that is carried 
beyond the leading order in perturbation theory.  However, 
Tevatron and Large Hadron Collider (LHC) experiments measure 
{\em isolated} photons in which part of the real-emission final-state phase 
space is excluded, resulting in incomplete cancellation of singularities in 
an analytic calculation~\cite{Berger:1995cc}.  
Isolation is required since, otherwise, the prompt photon signal would be 
overwhelmed by secondary $\gamma$'s from hadron decays, and it is hard 
to discern a prompt $\gamma$ if it is buried in a jet (unless $z$ is 
very large). In isolation procedures applied by the experiments, the 
energy of the hadronic remnants is required to be less than a 
threshold isolation energy 
$E_{T}^{iso}$ in a cone $\Delta R=\sqrt{\Delta\eta^{2}+\Delta\varphi^{2}}$
around each photon, with $\Delta\eta$ and $\Delta\varphi$ being
the separations of the hadronic remnant(s) from the photon in the
plane of pseudo-rapidity $\eta$ and azimuthal angle $\varphi$.  
The values of $E_{T}^{iso}$ and $\Delta R$ characterize the measurement, 
and, in a theory calculation, 
the size of the fragmentation contributions depends on the assumed values 
of these parameters\footnote{ 
In the case of production of photon pairs, discussed below, the
two photons must also be separated in the $\eta-\varphi$ plane by
an amount exceeding the approximate angular size $\Delta R_{\gamma\gamma}$
of one calorimeter cell.}.  Theoretical representations of isolation are at 
best approximate.  Calculations tend to require Monte Carlo modeling 
of fragmentation in order to implement experiment-like isolation selections.  

Calculations exist of prompt single photon production at next-to-leading order 
in QCD for both the direct and the 
fragmentation contributions.  One approach is encoded in the Monte Carlo 
program JETPHOX~\cite{Aurenche:2006vj}.  An impressive comparison with 
experiment may be found in the D0 paper~\cite{DZero}.  

Much more can be written about recent research on inclusive and 
isolated single photon production, but I turn to production of photon 
pairs in the remainder of this brief report, summarizing a new 
calculation of the diphoton cross section in perturbative 
QCD~\cite{Balazs:2006cc}. The motivation for this study comes, in 
part, from the fact that a Higgs boson with mass between 115 and 140~GeV 
may be identified at 
hadron colliders through its decay into a pair of energetic photons.  
Theoretical predictions of the QCD continuum `background' may be of 
substantial value in aiding search strategies.  Moreover, the 
QCD calculation of continuum photon-pair 
production is of theoretical interest in its own right, and data from the 
Tevatron collider~\cite{Acosta:2004sn} offer an opportunity to test results 
against experiment. 

I focus on the transverse momentum ($Q_T$) distribution of 
a pair of photons from the QCD continuum and from the Higgs boson signal.   
This distribution is important because its behavior affects the precision 
of the determination of the vertex from which the Higgs boson emerges. 
Greater activity in $Q_T$ associated with Higgs boson production allows a 
more precise determination of the event vertex.  The expected shape of 
$d\sigma/ d Q_T$ can affect triggering and analysis strategies.  Event 
modeling, kinematic  acceptance, and efficiencies all depend on $Q_T$.  
Finally, as is shown here, selections on $Q_T$ may be useful to enhance 
the signal to continuum ratio.  

\section {Calculation of the Diphoton Continuum} 

Direct production of continuum photon pairs occurs from $q\bar{q}$,
$q^{^{\!\!\!\!\!\!\!(-)}}g,$ and $gg$ scattering. In 
Ref.~\cite{Balazs:2006cc},  next-to-leading order (NLO) perturbative 
cross sections are included, i.e., cross sections of 
${\mathcal{O}}(\alpha_{s})$ in the $q\bar{q}$ and $qg$ channels, 
and ${\mathcal{O}}(\alpha_{s}^{3})$ in the $gg$ channel.  
Singular logarithms arise in the NLO cross sections when $Q_{T}$ of
the $\gamma\gamma$ pair is much smaller than its invariant mass $Q$.  
To describe properly the behavior of the $Q_T$ distribution in this 
region $Q_T < Q$, where the cross section is greatest, all-orders 
resummation of initial-state soft and collinear logarithmic contributions 
is included, based on the Collins-Soper-Sterman (CSS) resummation 
procedure~\cite{Collins:1984kg}.  The logarithms are (re)summed into 
a Sudakov exponent (composed of functions $A(\mu)$ and $B(\mu)$) and 
convolutions of the conventional parton densities $f_{a}(x,\mu_{F})$ 
with Wilson coefficient functions $C$. We include the 
${\mathcal{O}}(\alpha_{s}^{3})$ expressions for $A(\mu)$, 
${\mathcal{O}}(\alpha_{s}^{2})$ 
expressions for $B(\mu)$, and the $C$-functions at order $\alpha_{s}$, 
for all subprocesses.  
These enhancements elevate the accuracy of the resummed prediction
to the next-to-next-to-leading-logarithmic (NNLL) level. 

\section {Comparison with Tevatron Data} 
\begin{figure}
  \includegraphics[height=.3\textheight]{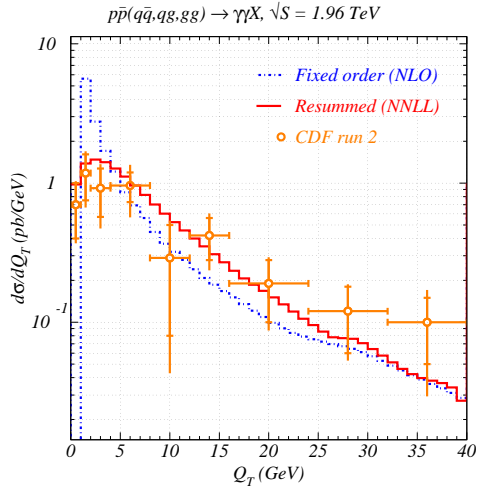}
\caption{Transverse momentum distributions of continuum diphotons for  
$p\bar{p} \rightarrow \gamma \gamma X$ at $\sqrt{s}=1.96$~TeV.  The dotted 
histogram shows the result of the next-to-leading order calculation 
and the solid histogram the result after resummation at NNLL 
accuracy. The data are from the CDF measurement.}
\label{fig:TEVQT}
\end{figure}

Our analysis provides the triple-differential cross section 
$d \sigma/dQ dQ_T d \Delta \phi$ ($\Delta \phi$ is the difference of 
the angles of the two photons in the transverse plane).  The 
calculation is especially pertinent 
for the transverse momentum distribution in the region 
$Q_T \le Q$, for fixed values of diphoton mass $Q$.  It would 
be best to compare our multi-differential 
distribution with experiment, but the collider data 
tend to be presented in the form of single-differential 
distributions in $Q$, $Q_T$, and $\Delta \phi$, after integration 
over the other variables.  For calculations appropriate at Tevatron 
energies, we impose 
cuts $|y_{\gamma}|<0.9$ on the rapidity of each photon, and 
$p_{T}^{\gamma}>p_{Tmin}^{\gamma}= $~14 (13) GeV on the 
transverse momentum of the harder (softer) photon in each 
$\gamma\gamma$ pair.  We choose $E_{T}^{iso}=1$ GeV, 
$\Delta R=0.4,$ and $\Delta R_{\gamma\gamma}=0.3$.  

The $Q_{T}$ distribution is shown in Fig.~\ref{fig:TEVQT}.  The resummed 
$Q_T$ distribution is well-behaved as 
$Q_T \rightarrow 0$, unlike its fixed-order counterpart which is 
singular in this limit. The result after NNLL resummation is in good  
agreement with the absolute rate and overall shape of the data 
from the Collider Detector at Fermilab (CDF) collaboration at 
$p\bar{p}$ collision energy $\sqrt{s}=1.96$ TeV~\cite{Acosta:2004sn}.  

In the two highest-$Q_{T}$ bins, there is some evidence of an excess 
experimental rate, perhaps a {}`shoulder' in the data.  
Our calculations show that most of the shoulder events populate 
a limited volume of phase space characterized
by small $\Delta\phi$ and $Q_T > Q$.  From a theoretical point of view, when 
$Q_{T}>Q$, as in the shoulder region, the calculation must be 
organized in a different way~\cite{Berger:1998ev} in order
to resum contributions arising from the fragmentation of partons into a 
pair of photons with small invariant mass, a new feature of diphoton 
production with respect to single photon production.  
Adequate treatment of the light $\gamma\gamma$ pairs 
is missing in our calculation and in all other calculations at present.
This interesting region warrants further theoretical investigation.
Our theoretical treatment is most reliable 
in the region $Q_{T}<Q$.  When this selection is made, the 
contributions from small $\Delta\phi$  are
efficiently suppressed, and dependence on tunable isolation parameters
and factorization scales is reduced.  After the  
selection $Q_T < Q$, we expect that the large $Q_T$ shoulder will disappear 
in the experimental $Q_T$  distribution.

Predicted distributions in the invariant mass $Q$ of the photon 
pair and the separation $\Delta \phi$ may be found in 
Ref.~\cite{Balazs:2006cc}.  Suggestions are made there for more differential 
analyses of the Tevatron data that would allow refined tests of the calculation.
An important consequence of the resummation formalism is a logarithmic dependence 
on the diphoton invariant mass $Q$.  We encourage the CDF and D0 collaborations to 
verify this predicted broadening of the $Q_T$ distribution as $Q$ increases. 

\section{Predictions for the LHC and comparison with the Higgs boson signal}
\begin{figure}
  \includegraphics[height=.3\textheight]{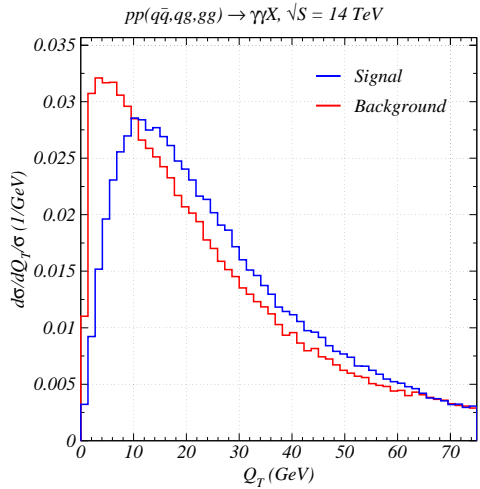}
\caption{Comparison of the normalized transverse momentum distributions  
for the Higgs boson signal and diphoton continuum at the LHC, 
both computed at NNLL accuracy.  The continuum is 
calculated for $128 < Q < 132$~GeV, and the Higgs boson mass 
is taken to be $m_H = 130$~GeV.}
\label{fig:HIGGSQT}
\end{figure}
Several predictions are presented in Ref.~\cite{Balazs:2006cc} for the 
photon-pair continuum in $p p$ collisions at $\sqrt s = 14$~TeV.  In 
this summary, I limit my discussion to differences 
expected between the spectrum of photon pairs from the 
Higgs boson signal and from the QCD continuum.  The  
dominant production mechanism for the Higgs boson is gluon fusion, 
$gg \to H \to \gamma\gamma$, and we calculate 
the signal with the same order of precision in the QCD contributions as for the 
continuum.  We include fixed-order initial state QCD corrections at $O(\alpha_s^3)$ 
(NLO) and resummation at NNLL accuracy.  These contributions are coded in the same 
numerical program used to compute the continuum, and we apply the same cuts (for 
example, on the values of $p_T$ and $\eta$ of each photon) for both 
the signal and the continuum.  Use of one code minimizes potential deviations than 
can arise from codes with different physics content, parton 
densities, and numerical implementations. We compute the continuum in the range 
$128 < Q < 132$~GeV and the signal at a fixed Higgs boson mass $m_H = 130$~GeV.  
We use $\mu_R = \mu_F = Q$ for the signal and the continuum.

The cross section times branching ratio for the Higgs boson signal is 
substantially smaller than the continuum rate.   
We present distributions normalized to the respective total 
rates in Fig.~\ref{fig:HIGGSQT}.  The signal and continuum peak at about 
12 and 5~GeV, respectively.  The dominant Sudakov exponent 
$C_F = A^{(1)}_{q\bar{q}} = A^{(1)}_{qg}$ that controls gluon radiation  
for the $q \bar{q}$ and $qg + \bar{q}g$ initial states is less than 
$A^{(1)}_{gg} = C_A$ that controls the $gg$ case.  More gluon radiation 
in the $gg$ case explains the broader $Q_T$ spectrum for the Higgs 
boson signal. The comparison in Fig.~\ref{fig:HIGGSQT} suggests 
that the signal to background ratio would be enhanced if a cut is made to 
restrict $Q_T > 10$~GeV.  Additional differences  between the signal and 
continuum are shown in Ref.~\cite{Balazs:2006cc}, including the rapidity 
difference of the two photons,
the scattering angle in the 
Collins-Soper frame $\tanh((y^{\gamma}_1- y^{\gamma}_2)/2) = \cos \theta^*$, 
and the difference between the azimuthal angles of the photons.   

\begin{theacknowledgments}
  
Work in the High Energy Physics Division at Argonne is supported by the 
U.~S.~Department of Energy, Division of High Energy Physics, Contract
W-31-109-ENG-38. 

\end{theacknowledgments}

\end{document}